\documentclass[final,5p,times,twocolumn]{elsarticle}
\usepackage{lineno,hyperref,amsmath,amsthm,amssymb,amsfonts,ragged2e,color,subfig}
\journal{``Contributions to Plasma Physics"}
\begin{document}
\begin{frontmatter}
\title{Modulational instability of dust-ion-acoustic waves and associated  envelope solitons in a non-thermal plasma}
\author{M.K. Islam$^{*,1}$, B.E. Sharmin$^{**,1}$, S. Biswas$^{***,1}$, M. Hassan$^{\dag,1}$, A.A. Noman$^{\ddag,1}$,\\
 N.A. Chowdhury$^{\S,2}$,  A. Mannan$^{\ddag\ddag,1,3}$, and A.A. Mamun$^{\S\S,1}$}
\address{$^{1}$Department of Physics, Jahangirnagar University, Savar, Dhaka-1342, Bangladesh\\
$^2$ Plasma Physics Division, Atomic Energy Centre, Dhaka-1000, Bangladesh\\
$^3$ Institut f\"{u}r Mathematik, Martin Luther Universit\"{a}t Halle-Wittenberg, Halle, Germany\\
e-mail: $^*$islam1243phy@gmail.com, $^{**}$sharmin146phy@gmail.com, $^{***}$shawonbiswas440@gmail.com,\\
$^{\dag}$hassan206phy@gmail.com, $^{\ddag}$noman179physics@gmail.com, $^{\S}$nurealam1743phy@gmail.com,\\
$^{\ddag\ddag}$abdulmannan@juniv.edu, $^{\S\S}$mamun\_phys@juniv.edu}
\begin{abstract}
A theoretical investigation has been made to understand the mechanism of the formation of both bright
and dark envelope soltions associated with dust-ion-acoustic waves (DIAWs) propagating
in an unmagnetized three component dusty plasma medium having inertial warm  positive ions and negative
dust grains, and inertialess non-thermal Cairns' distributed electrons. A nonlinear Schr\"{o}dinger equation (NLSE)
is derived by employing reductive perturbation method. The effects of
plasma parameters, viz.,  $\gamma_2$ (the ratio of the positive ion temperature  to
electron temperature times the charge state of ion) and $\nu$ (the ratio of the charge state of negative dust grain to positive ion)
on the modulational instability of DIAW which is governed by NLSE, are extensively studied. It is found that increasing
the value of the ion (electron) temperature reduces (enhances) the critical wave number ($k_c$).
The results of our present theoretical work may be used to interpret the nonlinear electrostatic structures which can exist
in many astrophysical environments and laboratory plasmas.
\end{abstract}
\begin{keyword}
Dust-ion-acoustic waves \sep NLSE \sep Modulational instability \sep  Envelope solitons.
\end{keyword}
\end{frontmatter}
\section{Introduction}
\label{2sec:Introduction}
A dusty plasma (DP), which is defined as fully or partially ionized electrically conducting
low-temperature gas, is referred as ``complex plasma" due to the existence of the micron or
sub-micron sized dust grains \cite{Shukla2002,Shahmansouri2014,Alinejad2010,Paul2016,Banerjee2016}.
The presence of massive dust grains significantly modifies the dynamics
of the DP medium (DPM) \cite{Sardar2017,Amin1998,El-Labany2015,Saini2008,Misra2006}.
The size and shape of the dust grains (million times heavier than the protons and their sizes range from
nanometres to millimetres) are considerable with those of the ions/protons \cite{Shukla2002,Shahmansouri2014,Alinejad2010}.
Over the last few decades, there has been a great interest in investigating the linear and nonlinear
wave propagation in DPM which can be found in both space environments (viz.,  cometary tails \cite{Alinejad2010},
the magnetosphere of the Jupiter and the Saturn \cite{Paul2016}, interstellar medium \cite{Banerjee2016,Sardar2017},
in the galactic centre \cite{Sardar2017}, and the Earth's ionosphere \cite{Saini2008}, etc.) and also laboratory plasmas (viz., electronics industry \cite{Selwyn1993,Kersten2001}). For the generation and propagation of electrostatic waves in DPM,
the moment of inertia is basically contributed by the heavy elements of the medium while the restoring force is contributed
by the light elements. The moment of inertia (restoring force) is contributed by the mass of the massive dust grains
(thermal pressure of the electrons and ions) for the formation of the dust-acoustic waves (DAWs) \cite{Amin1998,El-Labany2015,Saini2008,Misra2006} in three components DPM (viz., electrons, ions, and dust grains, etc). On the other hand, in the propagation of the dust-ion-acoustic waves (DIAWs) \cite{Alinejad2010,Paul2016,Banerjee2016,Sardar2017}, the moment of inertia (restoring force) is
contributed by the mass of the massive ions (thermal pressure of the electrons) in the presence of immobile
massive dust grains. So, the massive  dust grains do not play  direct role in the formation of DIAWs but their existence in the
background rigorously changes the dynamics of the DPM.

The existence of non-Maxwellian particles has been common in most of the space and laboratory DPM,
and thus many scientists have been interested to analyse the behaviour of nonlinear electrostatic potential structures
in the non-thermal DPM. Vela satellite has been observed that the electrons and ions in the Earth's bow-shock do
follow non-Maxwellian velocity distribution \cite{Hundhausen1967} instead of Maxwellian velocity distribution.
Cairns' \textit{et al.} \cite{Cairns1995} first constructed the non-thermal velocity distribution function
for explaining the nonlinear behaviour of the space plasma species such as electrons and ions, and successively, this
distribution has been considered by many authors \cite{Alinejad2010,Banerjee2016,Amin1998} for further treatment
to non-thermal plasma species. Alinejad \cite{Alinejad2010} studied DIAWs in a non-thermal DPM having inertial
ions and inertialess electrons in the presence of immobile massive dust grains, and observed that the width of
electrostatic pulse increases with electrons non-thermality. Paul and Bandyopadhyay \cite{Paul2016}
investigated the nonlinear properties of DIAWs in DPM by considering inerialess non-thermal Cairns'
distributed electrons and inertial ions as well as  massive dust grains in the background. Banerjee and
Maitra \cite{Banerjee2016} examined the condition for the formation of positive
potential solitary waves with different values of non-thermal parameter $\alpha$
in a multi-component DPM.

Bright and dark envelope solitions can generate due to the existence of external perturbations in
a nonlinear dispersive medium, and are considered two important solitonic solutions of the standard nonlinear
Schr\"{o}dinger equation (NLSE) which governs the modulational instability (MI) of the carrier waves.
Amin \textit{et al.} \cite{Amin1998} studied the MI of the DAWs and DIAWs in a three component DPM.
El-Labany \textit{et al.} \cite{El-Labany2015} theoretically and numerically analyzed the instability criteria
of the DAWs in the presence of non-thermal plasma species. Misra and Chowdhury \cite{Misra2006} considered
inertial massive dust grains and inertialess electrons and ions for studying the MI of DAWs. To the best
knowledge of authors, no one has considered inertial ions along with inertial dust grains
and inertialess non-thermal electrons to investigate DIAWs and associated MI of DIAWs. Hence, in
this paper, we would like to investigate the MI of the DIAWs  in which the moment of inertia is provided
by the mass of the inertial negatively charged dust grains as well as warm ions, and the restoring force
is provided by the thermal pressure of the non-thermal electrons.

The rest part of this paper goes as follows: The governing equations are presented
in section \ref{2sec:Governing Equations}. The derivation of NLSE via reductive
perturbation method (RPM) is demonstrated in section \ref{2sec:Derivation of the NLSE}.
The MI and envelope solitons are provided in section \ref{2sec:Modulational instability and Envelope Solitons}.
Results and discussions are provided in section \ref{2sec:Result and discussions}.
A brief conclusion is provided in section \ref{2sec:Conclusion}.
\section{Governing equations}
\label{2sec:Governing Equations}
We consider a three component DPM comprising of inertial positively charged warm ions (charge $q_i=Z_ie$
and mass $m_i$) and inertial negatively charged dust grains (charge $q_d=-Z_de$ and mass $m_d$) as
well as inertialess non-thermal electrons (charge $q_e=-e$; mass $m_e$);
where $Z_i$ ($Z_d$) is the number of  protons (electrons) residing on the ion (dust grain)
surface, and $e$ is the magnitude of the charge of an electron.
Overall, the charge neutrality condition for our plasma model is written as $ Z_in_{i0} = Z_d n_{d0}+ n_{e0}$.  Now, the normalized governing equations
of the DIAWs can be written as
\begin{eqnarray}
&&\hspace*{-1.3cm}\frac{\partial n_{d}}{\partial t}+\frac{\partial}{\partial x}(n_{d}u_{d})=0,
\label{2eq:1}\\
&&\hspace*{-1.3cm}\frac{\partial u_{d}}{\partial t}+u_{d}\frac{\partial u_{d}}{\partial x}=\gamma_1\frac{\partial\phi}{\partial x},
\label{2eq:2}\\
&&\hspace*{-1.3cm}\frac{\partial n_{i}}{\partial t}+\frac{\partial}{\partial x}(n_{i}u_{i})=0,
\label{2eq:3}\\
&&\hspace*{-1.3cm}\frac{\partial u_{i}}{\partial t}+u_{i}\frac{\partial u_{i}}{\partial x}+\gamma_2 n_i\frac{\partial n_i}{\partial x}=-\frac{\partial\phi}{\partial x},
\label{2eq:4}\\
&&\hspace*{-1.3cm}\frac{\partial^2\phi}{\partial x^2}=\gamma_3 n_e+(1-\gamma_3)n_d-n_i,
\label{2eq:5}
\end{eqnarray}
where $n_d$ $(n_i)$ is the dust (ion) number density normalized by its equilibrium
value $n_{d0} $ $(n_{i0})$; $u_d$ $(u_i)$ is the dust (ion) fluid speed normalized by
the ion-acoustic wave speed $C_i=(Z_i k_BT_e/m_i)^{1/2}$ with $T_e$ being the non-thermal
electron temperature and $k_B$ being the Boltzmann constant; $\phi$ is the electrostatic wave potential normalized
by $k_BT_e/e$; the time and space variables are normalized by ${\omega^{-1}_{pi}}=(m_i/4\pi {Z_i}^2 e^2 n_{i0})^{1/2}$
and $\lambda_{Di}=(k_BT_e/4 \pi Z_i e^2 n_{i0})^{1/2}$, respectively. The pressure term of the ion is recognized as
$P_i=P_{i0}(N_i/n_{i0})^\gamma$ with $P_{i0}=n_{i0}k_BT_i$ being the equilibrium
pressure of the ion, $T_i$ being the temperature of warm ion, and
$\gamma=(N+2)/N$ (where $N$ is the degree of freedom and for one-dimensional case
$N=1$, hence $\gamma=3$). Other parameters can be defined as $\gamma_1=\mu\nu$,
$\mu=m_i/m_d$, $\nu=Z_d/Z_i$, $\gamma_2=3T_i/Z_i T_e$, and $\gamma_3=n_{e0}/Z_i n_{i0}$.
Now, the expression for  electron number density which is obeying non-thermal Cairns' distribution
\cite{Cairns1995} is given by
\begin{eqnarray}
&&\hspace*{-1.3cm}n_e=(1-\beta\phi+\beta\phi^2)\exp(\phi),
\label{2eq:6}
\end{eqnarray}
where $\beta=4\alpha/(1+3\alpha)$ with $\alpha$ being the parameter determining the faster particles present in plasma
model. Now, by substituting Eq. \eqref{2eq:6} into Eq. \eqref{2eq:5}, and expanding the exponential term up to third order, we can find
\begin{eqnarray}
&&\hspace*{-1.3cm}\frac{\partial^2\phi}{\partial x^2}+n_i=\gamma_3+(1-\gamma_3)n_d+H_1\phi+H_2\phi^2+H_3\phi^3+\cdot\cdot\cdot,
\label{2eq:7}\
\end{eqnarray}
where $H_1 = \gamma_3-\gamma_3\beta$, $H_2 = \gamma_3/2$, and $H_3 = \gamma_3/6-\gamma_3\beta/2$.
The terms $H_1$, $H_2$, and $H_3$ in the right-hand side
of Eq. \eqref{2eq:7} are the contribution of inertialess electrons.
\section{Derivation of the NLSE}
\label{2sec:Derivation of the NLSE}
In order to investigate the MI and envelope solitons associated with DIAWs, we drive the NLSE by applying the
RPM. At first, we introduced the stretched co-ordinates in the following form \cite{C1,C2,C3,C4,C5,C6,C7}:
\begin{eqnarray}
&&\hspace*{-1.3cm}\xi={\epsilon}(x-v_g t),~~~\tau={\epsilon}^2 t,
\label{2eq:8}
\end{eqnarray}
where $v_g$ denotes the group speed  of the carrier waves and $\epsilon$ represents  nonlinear  parameter.
The dependent variables can be written as \cite{C7,C8,C9,C10,C11,C12,C13}
\begin{eqnarray}
&&\hspace*{-1.3cm} n_{dl}^m=1+\sum_{m=1}^{\infty} \epsilon^{m} \sum_{l=-\infty}^{\infty} n_{dl}^{(m)} (\xi,\tau) \exp[il(kx-\omega t)],
\label{2eq:9}\\
&&\hspace*{-1.3cm} u_{dl}^m=\sum_{m=1}^{\infty} \epsilon^{m} \sum_{l=-\infty}^{\infty} u_{dl}^{(m)} (\xi,\tau) \exp[il(kx-\omega t)],
\label{2eq:10}\\
&&\hspace*{-1.3cm} n_{il}^m=1+\sum_{m=1}^{\infty} \epsilon^{m} \sum_{l=-\infty}^{\infty} n_{il}^{(m)} (\xi,\tau) \exp[il(kx-\omega t)],
\label{2eq:11}\\
&&\hspace*{-1.3cm} u_{il}^m=\sum_{m=1}^{\infty} \epsilon^{m} \sum_{l=-\infty}^{\infty} u_{il}^{(m)} (\xi,\tau) \exp[il(kx-\omega t)],
\label{2eq:12}\\
&&\hspace*{-1.3cm}\phi_l^m=\sum_{m=1}^{\infty} \epsilon^{m} \sum_{l=-\infty}^{\infty} \phi_{l}^{(m)} (\xi,\tau) \exp[il(kx-\omega t)],
\label{2eq:13}\
\end{eqnarray}
where $k$ ($\omega$) indicates the carrier wave number (frequency).
We can represent the derivative operators as
\begin{eqnarray}
&&\hspace*{-1.3cm}\frac{\partial}{\partial t}\rightarrow\frac{\partial}{\partial t}-\epsilon v_g \frac{\partial}{\partial \xi}+ \epsilon^2\frac{\partial}
{\partial \tau},
\label{2eq:14}\\
&&\hspace*{-1.3cm}\frac{\partial}{\partial x}\rightarrow\frac{\partial}{\partial x}+\epsilon\frac{\partial}{\partial \xi}.
\label{2eq:15}\
\end{eqnarray}
Now, by substituting Eqs. \eqref{2eq:8}-\eqref{2eq:15} into Eqs. \eqref{2eq:1}-\eqref{2eq:4} and Eq. \eqref{2eq:7}, and picking up
the terms that are associated with $\epsilon$, the first order ($m=1$ with $l=1$)
equations may be written as
\begin{eqnarray}
&&\hspace*{-1.3cm}n_{d1}^{(1)}=-\frac{\gamma_1k^2}{\omega^2}\phi_1^{(1)},
\label{2eq:16}\\
&&\hspace*{-1.3cm}u_{d1}^{(1)}=-\frac{\gamma_1 k }{\omega}\phi_1^{(1)},
\label{2eq:17}\\
&&\hspace*{-1.3cm}n_{i1}^{(1)}=\frac{k^2}{\omega^2-\gamma_2 k^2}\phi_1^{(1)},
\label{2eq:18}\\
&&\hspace*{-1.3cm}u_{i1}^{(1)}=\frac{k\omega}{\omega^2-\gamma_2 k^2}\phi_1^{(1)},
\label{2eq:19}\
\end{eqnarray}
and the dispersion relation for DIAWs can be written as
\begin{eqnarray}
&&\hspace*{-1.3cm}\omega^2=\frac{k^2B\pm k^2\sqrt{B^2-4AC}}{2A},
\label{2eq:20}\
\end{eqnarray}
\begin{figure}[t!]
\centering
\includegraphics[width=80mm]{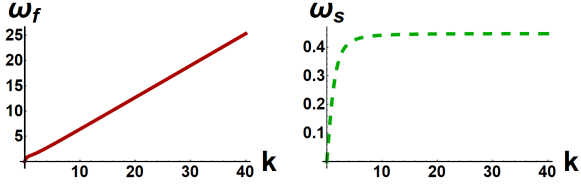}
\caption{Plot of $\omega_f$ vs $k$ (left panel) and  $\omega_s$ vs $k$ (right panel)
when $\alpha=0.4$, $\mu=10^{-6}$, $\nu=5\times10^5$, $\gamma_2=0.4$, and $\gamma_3=0.6$.}
\label{2Fig:F1}
\end{figure}
where $A=k^2+H_1$, $B=1+\gamma_1-\gamma_1\gamma_3+\gamma_2 k^2+\gamma_2 H_1$, and $C=\gamma_1\gamma_2-\gamma_1\gamma_2\gamma_3$.
In Eq. \eqref{2eq:20}, to get real and positive value of $\omega$, the condition $B^2>4AC$ should be
satisfied. The positive and negative signs in Eq. \eqref{2eq:20} correspond to the fast ($\omega_f$) and slow ($\omega_s$) DIA modes.
The fast DIA mode corresponds to the case in which both inertial dust and ion components oscillate in phase with the inertialess
electrons. On the other hand, the slow DIA mode corresponds to the case in which only one of the inertial components
oscillates in phase with inertialess electrons, but the other inertial component oscillates in anti-phase with
them \cite{Dubinov2009,Saberiana2017}. We have numerically analyzed the fast and slow DIA modes in Fig. \ref{2Fig:F1}
in the presence of non-thermal electrons. Figure  \ref{2Fig:F1} (left panel) indicates
that the frequency of the DIAWs can be higher than ion-plasma frequency. In this regard, we note that in absence of dust,
the frequency of the ion-acoustic waves is always less than the ion-plasma or ion-Langmuir frequency. However,
the phase speed of the DIAWs  increases with the magnitude of the dust charge ($Z_d$) and dust number
density ($n_{d0}$). This is due to the extra space charge electric field created by the highly negatively
charged dust grains. This is theoretically predicted by Shukla and Silin \cite{Shukla1992} and experimentally
observed by Barkan \textit{et al.} \cite{Barkan1996}. Thus, as the magnitude of the dust charge ($Z_d$) or
dust number density ($n_{d0}$) increases, the frequency  of the DIAWs increases, even it can exceed the
ion-plasma or ion-Langmuir frequency. On the other hand, the dispersion curve of slow DIA mode shown in
Fig. \ref{2Fig:F1} (right panel) clearly indicates that the frequency of the slow DIA mode is always less
than the ion-plasma or ion-Langmuir frequency even in the presence of highly negatively charged dust.

For second order harmonics, equations can be found from the next order of $\epsilon$ (with $m=2$ and $l=1$) as
\begin{eqnarray}
&&\hspace*{-1.3cm}n_{d1}^{(2)}=-\frac{\gamma_1 k^2}{\omega^2}\phi_1^{(2)}-\bigg[\frac{2(\gamma_1 k\omega-\gamma_1 v_g k^2)}{i\omega^3}\bigg] \frac{\partial \phi_1^{(1)}}{\partial\xi},
\label{2eq:21}\\
&&\hspace*{-1.3cm}u_{d1}^{(2)}=-\frac{\gamma_1 k}{ \omega}\phi_1^{(2)} -\bigg[\frac{\gamma_1\omega-\gamma_1 v_gk}{i\omega^2}\bigg] \frac{\partial \phi_1^{(1)}}{\partial\xi},
\label{2eq:22}\\
&&\hspace*{-1.3cm}n_{i1}^{(2)}=\frac{k^2}{\omega^2-\gamma_2 k^2}\phi_1^{(2)} +\bigg[\frac{2k\omega(\omega-v_gk)}{i(\omega^2-\gamma_2 k^2)^2}\bigg] \frac{\partial \phi_1^{(1)}}{\partial\xi},
\label{2eq:23}\\
&&\hspace*{-1.3cm}u_{i1}^{(2)}=\frac{k\omega}{\omega^2-\gamma_2 k^2}\phi_1^{(2)}+\bigg[\frac{(\omega-v_gk)}{i(\omega^2-\gamma_2 k^2)^2}\bigg]\frac{\partial \phi_1^{(1)}}{\partial\xi}
\nonumber\\
&&\hspace*{-0.5cm}+\bigg[\frac{(\omega^2+\gamma_2k^2)}{i(\omega^2-\gamma_2 k^2)^2}\bigg]\frac{\partial \phi_1^{(1)}}{\partial\xi},
\label{2eq:24}\
\end{eqnarray}
with the compatibility condition, we have obtained the group speed of IAWs as
\begin{eqnarray}
&&\hspace*{-1.3cm}v_g=\frac{\gamma_1\gamma_3\omega^5+2\gamma_1\gamma_2 k^2\omega^3-2\gamma_1\gamma_2\gamma_3 k^2\omega^3-\gamma_1\omega^5+\mathcal{F}_1}{\gamma_1\gamma_2^2\gamma_3k^5-2\gamma_1\gamma_2\gamma_3 k^3\omega^2+\mathcal{F}_2},
\label{2eq:25}
\end{eqnarray}
where
\begin{eqnarray}
&&\hspace*{-1.3cm}\mathcal{F}_1=\gamma_1\gamma_2^2\gamma_3 k^4\omega-\omega^5+\omega^7-2\gamma_2\omega^5 k^2+\gamma_2^2\omega^3 k^4-\gamma_1\gamma_2^2 k^4\omega,
\nonumber\\
&&\hspace*{-1.3cm} \mathcal{F}_2=\gamma_1\gamma_3k\omega^4-\gamma_1k\omega^4+2\gamma_1\gamma_2 k^3\omega^2-k\omega^4-\gamma_1\gamma_2^2k^5.
\nonumber\
\end{eqnarray}
The coefficients of the $\epsilon$ when  $m=2$ with $l=2$ provides the second
order harmonic amplitudes which are found to be proportional to $|\phi_1^{(1)}|^2$
\begin{eqnarray}
&&\hspace*{-1.3cm}n_{d2}^{(2)}=H_4|\phi_1^{(1)}|^2,
\label{2eq:26}\\
&&\hspace*{-1.3cm}u_{d2}^{(2)}=H_5 |\phi_1^{(1)}|^2,
\label{2eq:27}\\
&&\hspace*{-1.3cm}n_{i2}^{(2)}=H_6|\phi_1^{(1)}|^2,
\label{2eq:28}\\
&&\hspace*{-1.3cm}u_{i2}^{(2)}=H_7 |\phi_1^{(1)}|^2,
\label{2eq:29}\\
&&\hspace*{-1.3cm}\phi_{2}^{(2)}=H_8 |\phi_1^{(1)}|^2,
\label{2eq:30}\
\end{eqnarray}
where
\begin{eqnarray}
&&\hspace*{-1.3cm}H_4=\frac{3\gamma_1^2 k^4-2\gamma_1 k^2\omega^2H_8}{2\omega^4},
\nonumber\\
&&\hspace*{-1.3cm}H_5=\frac{\gamma_1^2 k^3-2\gamma_1k\omega^2H_8}{2\omega^3},
\nonumber\\
&&\hspace*{-1.3cm}H_6=\frac{2k^2 H_8(\omega^2-\gamma_2 k^2)^2+k^4(\gamma_2 k^2+3\omega^2)}{2(\omega^2-\gamma_2 k^2)^3},
\nonumber\\
&&\hspace*{-1.3cm}H_7=\frac{2k(H_8+\gamma_2H_6)(\omega^2-\gamma_2k^2)^2+k^3\omega^2+\gamma_2 k^5}{2\omega(\omega^2-\gamma_2k^2)^2},
\nonumber\\
&&\hspace*{-1.3cm}H_8=\{k^4\omega^4(3\omega^2+\gamma_2 k^2)-3\gamma_1^2 k^4(1-\gamma_3)(\omega^2-\gamma_2 k^2)^3
\nonumber\\
&&\hspace*{-0.5cm}-2\omega^4 H_2(\omega^2-\gamma_2 k^2)^3\}/6k^2\omega^4(\omega^2-\gamma_2 k^2)^3.
\nonumber\
\end{eqnarray}
Again, when ($m=3$ with $l=0$) and ($m=2$ with $l=0$), we find these relations
\begin{eqnarray}
&&\hspace*{-1.3cm}n_{d0}^{(2)}=H_9|\phi_1^{(1)}|^2,
\label{2eq:31}\\
&&\hspace*{-1.3cm}u_{d0}^{(2)}=H_{10}|\phi_1^{(1)}|^2,
\label{2eq:32}\\
&&\hspace*{-1.3cm}n_{i0}^{(2)}=H_{11}|\phi_1^{(1)}|^2,
\label{2eq:33}\\
&&\hspace*{-1.3cm}u_{i0}^{(2)}=H_{12}|\phi_1^{(1)}|^2,
\label{2eq:34}\\
&&\hspace*{-1.3cm}\phi_0^{(2)}=H_{13} |\phi_1^{(1)}|^2,
\label{2eq:35}\
\end{eqnarray}
where
\begin{eqnarray}
&&\hspace*{-1.3cm}H_{9}=\frac{2\gamma_1^2 v_g k^3+\gamma_1^2k^2\omega-\gamma_1\omega^3H_{13}}{v_g^2\omega^3},
\nonumber\\
&&\hspace*{-1.3cm}H_{10}=\frac{\gamma_1^2k^2-\gamma_1\omega^2H_{13}}{v_g\omega^2},
\nonumber\\
&&\hspace*{-1.3cm}H_{11}=\frac{H_{13}(\omega^2-\gamma_2k^2)^2+2 v_gk^3\omega+\gamma_2k^4+k^2\omega^2}{(\gamma_2-v_g^2)(\omega^2-\gamma_2k^2)^2},
\nonumber\\
&&\hspace*{-1.3cm}H_{12}=\frac{(H_13+\gamma_2H_{11})(\omega^2-\gamma_2k^2)^2+k^2\omega^2+\gamma_2k^4}{v_g(\omega^2-\gamma_2k^2)^2},
\nonumber\\
&&\hspace*{-1.3cm}H_{13}= \frac{v_g^2 k^2\omega^3(\gamma_2k^2+\omega^2+2 v_g k\omega)-\mathcal{F}_3}{\omega^3(\omega^2-\gamma_2k^2)^2\times\mathcal{F}_4},
\nonumber\
\end{eqnarray}
where
\begin{eqnarray}
&&\hspace*{-1.4cm}\mathcal{F}_3=(1-\gamma_3)(v_g^2-\gamma_2)(\omega^2-\gamma_2k^2)^2(\gamma_1^2k^2\omega+2\gamma_1^2 v_g k^3)
\nonumber\\
&&\hspace*{-0.7cm}+2v_g^2\omega^3H_2(v_g^2-\gamma_2)(\omega^2-\gamma_2k^2)^2,
\nonumber\\
&&\hspace*{-1.4cm}\mathcal{F}_4=v_g^2H_1(v_g^2-\gamma_2)-v_g^2-\gamma_1(1-\gamma_3)(v_g^2-\gamma_2).
\nonumber\
\end{eqnarray}
Now, we develop the standard NLSE by substituting all the above equations into third order harmonic modes ($m=3$ with $l=1$):
\begin{eqnarray}
&&\hspace*{-1.3cm}i\frac{\partial\Phi}{\partial\tau}+P\frac{\partial^2\Phi}{\partial\xi^2}+Q|\Phi|^2\Phi=0,
\label{2eq:36}
\end{eqnarray}
where $\Phi=\phi_1^{(1)}$ for simplicity. In Eq. \eqref{2eq:36}, $P$ can be written as
\begin{eqnarray}
&&\hspace*{-1.3cm}P=\frac{\mathcal{F}_5-\omega^4(\omega^2-\gamma_2 k^2)^3}{\omega (\omega^2-\gamma_2 k^2)\times\mathcal{F}_6},
\nonumber\
\end{eqnarray}
where
\begin{eqnarray}
&&\hspace*{-1.3cm}\mathcal{F}_5=\omega^4(\omega-v_gk)\{2 k\omega(\gamma_2 k-v_g\omega)+(\omega-v_g k)(\omega^2+\gamma_2 k^2)\}
\nonumber\\
&&\hspace*{-0.5cm}+\gamma_1(1-\gamma_3)(\omega-v_g k)(\omega-3v_gk)(\omega^2-\gamma_2 k^2)^3,
\nonumber\\
&&\hspace*{-1.3cm}\mathcal{F}_6=2 k^2\omega^4+2\gamma_1 k^2(1-\gamma_3)(\omega^2-\gamma_2 k^2)^2,
\nonumber\
\end{eqnarray}
and also $Q$ can be written as
\begin{eqnarray}
&&\hspace*{-1.3cm}Q=\frac{\mathcal{F}_7-\omega^3(\omega^2 k^2+\gamma_2 k^4)(H_6+H_{11})-2 k^3\omega^4(H_7+H_{12})}{\mathcal{F}_6},
\nonumber\
\end{eqnarray}
where
\begin{eqnarray}
&&\hspace*{-1.3cm}\mathcal{F}_7 =\big[\omega^3\{3H_3+2H_2(H_8+H_{13})\}-\gamma_1 k^2\omega(1-\gamma_3)(H_4+H_9)
\nonumber\\
&&\hspace*{-0.5cm}-2\gamma_1 k^3(1-\gamma_3)(H_5+H_{10})\big](\omega^2-\gamma_2 k^2)^2.
\nonumber\
\end{eqnarray}
The space and time evolution of the DIAWs in the plasma medium are directly governed by the  dispersion
($P$) and nonlinear  ($Q$) coefficients of NLSE and are indirectly governed by different plasma parameters such
as $\alpha$, $\mu$,  $\nu$, $\gamma_2$, and $\gamma_3$. Thus, these plasma parameters significantly
affect the stability conditions of DIAWs.
\begin{figure}[t!]
\centering
\includegraphics[width=70mm]{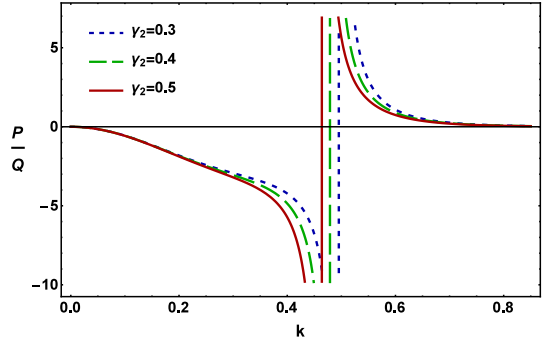}
\caption{Plot of $P/Q$ vs $k$ for various values of $\gamma_2$ when $\alpha=0.4$, $\mu=10^{-6}$, $\nu=5\times10^5$, and $\gamma_3=0.6$.}
\label{2Fig:F2}
\end{figure}
\begin{figure}[t!]
\centering
\includegraphics[width=70mm]{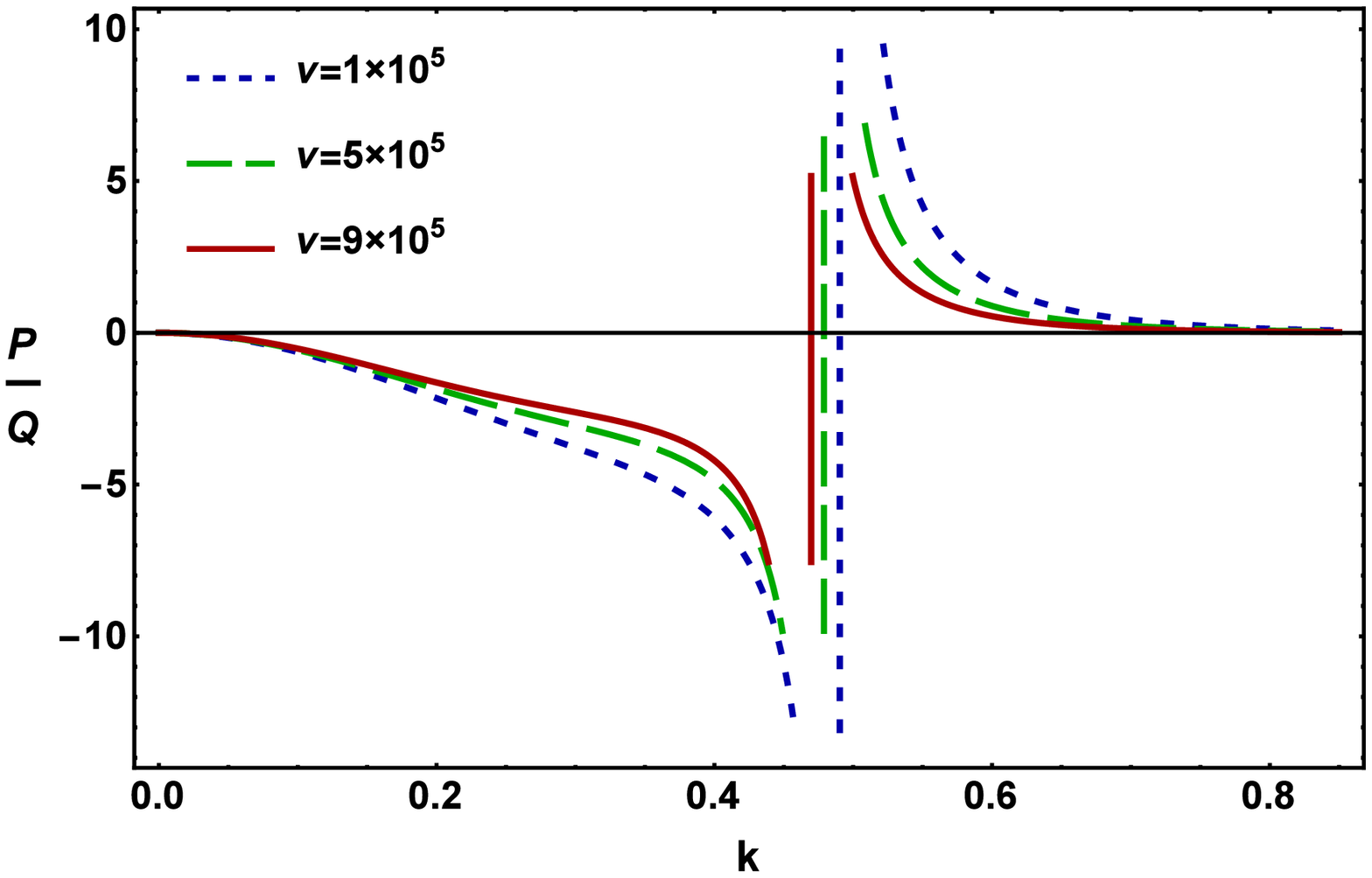}
\caption{Plot of $P/Q$ vs $k$ for various values of $\nu$ when $\alpha=0.4$, $\mu=10^{-6}$, $\gamma_2=0.4$, and $\gamma_3=0.6$.}
\label{2Fig:F3}
\end{figure}
\begin{figure}[t!]
\centering
\includegraphics[width=80mm]{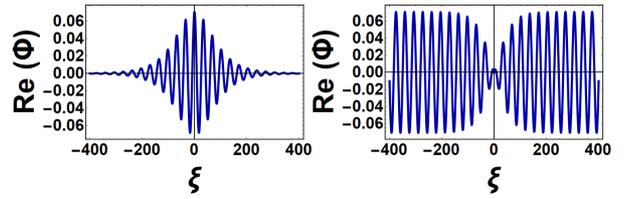}
\caption{Bright (left panel) and dark (right panel) envelope solitons  when other
plasma parameters are $\tau=0$, $\psi_0=0.005$, $U=0.4$, $\Omega_0=0.4$, $\alpha=0.4$, $\mu=10^{-6}$,
$\nu=5\times10^5$, $\gamma_2=0.4$, and $\gamma_3=0.6$.}
\label{2Fig:F4}
\end{figure}
\section{Modulational instability and Envelope Solitons}
\label{2sec:Modulational instability and Envelope Solitons}
The stable and unstable parametric regimes of DIAWs are
organised by the sign of $P$ and $Q$ of Eq. \eqref{2eq:36}.
When $P$ and $Q$ has the same sign (i.e., $P/Q > 0$),
the evolution of DIAWs amplitude is modulationally
unstable in the presence of external perturbations.
On the other hand, when $P$ and $Q$ has opposite sign
(i.e., $P/Q < 0$), the DIAWs are modulationally stable in
the presence of external perturbations. So, the plot of $P/Q$
against $k$ yields stable and unstable parametric regimes
of the DIAWs. The point at which the transition of $P/Q$
curve intersects with the $k$-axis is known as the critical wave number $k~(= k_c)$.

The bright (when $P/Q>0$) and dark (when $P/Q<0$) envelope
solitonic solutions can be written, respectively, as
\begin{eqnarray}
&&\hspace*{-1.3cm}\Phi(\xi,\tau)=\left[\psi_0~\mbox{sech}^2 \left(\frac{\xi-U\tau}{W}\right)\right]^\frac{1}{2}
\nonumber\\
&&\hspace*{-0.01cm}\times \exp \left[\frac{i}{2P}\left\{U\xi+\left(\Omega_0-\frac{U^2}{2}\right)\tau \right\}\right],
\label{2eq:37}\\
&&\hspace*{-1.3cm}\Phi(\xi,\tau)=\left[\psi_{0}~\mbox{tanh}^2 \left(\frac{\xi-U\tau}{W}\right)\right]^\frac{1}{2}
\nonumber\\
&&\hspace*{-0.01cm}\times \exp \left[\frac{i}{2P}\left\{U\xi-\left(\frac{U^2}{2}-2 P Q \psi_{0}\right)\tau \right\}\right],
\label{2eq:38}
\end{eqnarray}
where $\psi_0$ is the amplitude of localized pulse for both bright and dark
envelope solitons, $U$ is the propagation speed of the localized pulse, $W$ is the soliton width, and
$\Omega_0$ is the oscillating frequency at $U=0$. The soliton width $W$ and the maximum amplitude  $\psi_0$
are related as $W=\sqrt{2\mid P/Q\mid/\psi_0}$. We have depicted the bright (left panel) and dark (right panel)
envelope solitons in Fig. \ref{2Fig:F4}.
\section{Results and discussions}
\label{2sec:Result and discussions}
Now, we would like to numerically analyze the stability conditions of the DIAWs in the
presence of non-thermal electrons. The mass and charge state of
the plasma species, even their number density, are important factors in recognizing
the stability conditions of the DIAWs in DPM \cite{Shukla2002,Saberiana2017,Merlino2014,Shukla2012,Mamun2002,Shalaby2009}.
The mass of the dust grains is comparable to the mass of the protons. In a general picture
of the DPM, dust grains are massive (million to billion times heavier than the
protons) and their sizes range from nanometres to millimetres. Dust grains may be
metallic, conducting, or made of ice particulates. The size and shape of dust grains
will be different, unless they are man-made. The dust grains are million to billion
times heavier than the protons, and typically, a dust grain acquires one
thousand to several hundred thousand elementary charges \cite{Shukla2002,Saberiana2017,Merlino2014,Shukla2012,Mamun2002,Shalaby2009}.

It may be noted here that in DAWs, the mass of the dust grains provides
the moment of inertia, and the thermal pressure of the electrons and ions
provides the restoring force in a three component DPM. On the other hand, in DIAWs, the mass
of the ion provides the moment of inertia, and the thermal pressure of the electron
provides the restoring force in the presence of immobile dust grains.
In this article, we consider three component dusty plasma model having inertial warm
positive ions and negative dust grains, and interialess non-thermal electrons.
It may be noted here that in the DIAWs, if anyone considers the thermal effects of the ions
then it is important to consider the moment of inertia of the ions along with
the dust grains in the presence of inertialess electrons. This means that the consideration
of the pressure term of the ions highly contributes to the moment of inertia along with
inertial dust grains to generate DIAWs in a DPM having inertialess electrons.
In our present analysis, we have considered that $m_d=10^6m_i$, $Z_d=(10^3\thicksim10^5)Z_i$, and $T_e=10T_i$.

The effects of ion and electron temperature on MI conditions of DIAWs can be observed from Fig. \ref{2Fig:F2}
and it is clear from this figure that (a) it is really interesting that both modulaltionally stable and unstable
parametric regimes are allowed; (b) the DIAWs are modulationally stable for small values of $k$ while
modulationally unstable for large values of $k$; (c) the critical wave number $k_c$
decreases (increases) with increasing ion (electron) temperature for a constant value of $Z_i$ (via $\gamma_2$).
So, ion and electron temperature play an opposite role in recognizing the modulationally stable and unstable
parametric regimes of DIAWs. Figures \ref{2Fig:F3} can reflect the effects of the charge state of inertial warm ions and
negatively charged dust grains on the instability criterion of DIAWs in the presence of non-thermal electrons (via $\nu$).
The DIAWs become unstable for small (large) values of $k$ as we increase the charge state of the inertial
negatively charged dust grains (warm ions). Finally, from Fig. \ref{2Fig:F4}, it can be seen
that the bright (dark) envelope solitons associated with the unstable (stable) parametric regimes of DIAWs are allowed by the plasma model.
\section{Conclusion}
\label{2sec:Conclusion}
In this work, we have considered the moment of inertia of the warm ions along with negatively charged dust grains and inertialess non-thermal electrons
for studying the conditions of MI of the DIAWs. By employing RPM, we have derived the
NLSE \eqref{2eq:36} from  a set of basic equations, and have studied the formation of the electrostatic envelope solitons associated with DIAWs in an
unmagnnetized DPM. The consideration of the moment of inertia of the warm ions along with the negatively charged dust grains
in a three component DPM has significantly changed the dynamics of DPM as well as the instability conditions
of the DIAWs. We, finally, hope that the findings of our present investigation should be useful
in understanding the mechanism of the formation of electrostatic envelope solitons in a three
component DPM (viz., cometary tails \cite{Alinejad2010}, the magnetosphere of the Jupiter and
the Saturn \cite{Paul2016}, interstellar medium \cite{Banerjee2016,Sardar2017},
in the galactic centre \cite{Sardar2017}, and the Earth's ionosphere \cite{Saini2008}, etc.).

\end{document}